\documentstyle[aps,amsfonts,twocolumn,epsfig]{revtex}

\begin{document}
\draft

\title{Quantum fast Fourier transform using multilevel atoms}
\author{Ashok Muthukrishnan \cite{email} and C. R. Stroud, Jr.}
\address{The Institute of Optics, University of Rochester,
Rochester, New York 14627}
\date{August 27, 2001}
\maketitle

%**************************************************************
%
\renewcommand{\thesection}{\arabic{section}}
\renewcommand{\thesubsection}{\thesection.\arabic{subsection}}

\newcommand{\hs}[1]{\hspace{#1 ex}}
\newcommand{\vs}[1]{\vspace{#1 ex}}

\newcommand{\minus}{\hs{-0.2}-\hs{-0.2}}
\newcommand{\minusb}{\hs{-0.5}-\hs{-0.5}}
\newcommand{\equal}{\hs{-0.2}=\hs{-0.2}}
\newcommand{\equalb}{\hs{-0.5}=\hs{-0.5}}
\newcommand{\plus}{\hs{-0.2}+\hs{-0.2}}
\newcommand{\plusb}{\hs{-0.5}+\hs{-0.5}}
\newcommand{\gthan}{\hs{-0.2}>\hs{-0.2}}
\newcommand{\gthanb}{\hs{-0.5}>\hs{-0.5}}

\newcommand{\ket}[1]{\mbox{$|#1\rangle$}}
\newcommand{\ketn}[3]{\mbox{$|#1,#2,\ldots,#3\rangle$}}
\newcommand{\bra}[1]{\mbox{$\langle #1|$}}
\newcommand{\braket}[2]{\mbox{$\langle #1|#2 \rangle$}}
\newcommand{\ketbra}[2]{\mbox{$|#1\rangle\langle #2|$}}
\newcommand{\matrixelem}[3]{\mbox{$\langle #1|#2|#3 \rangle$}}

\newcommand{\cA}{\mbox{${\cal A}$}}
\newcommand{\cB}{\mbox{${\cal B}$}}
\newcommand{\cO}{\mbox{${\cal O}$}}

\newcommand{\nbar}{\bar{n}}

\newcommand{\Sch}{Schr\"odinger }

\newcommand{\postscript}[1] {\centerline{\epsfbox{#1}}}
%
%******************************************************************************

\widetext \hskip.5in
\begin{minipage}{5.75in}
\begin{abstract}
\vs{-7}

\hs{1.5} We propose an implementation of the quantum fast Fourier transform
algorithm in an entangled system of multilevel atoms. The Fourier transform
occurs naturally in the unitary time evolution of energy eigenstates and is
used to define an alternate wave-packet basis for quantum information in the
atom. A change of basis from energy levels to wave packets amounts to a
discrete quantum Fourier transform within each atom. The algorithm then reduces
to a series of conditional phase transforms between two entangled atoms in
mixed energy and wave-packet bases. We show how to implement such transforms
using wave-packet control of the internal states of the ions in the linear
ion-trap scheme for quantum computing.

%\pacs{PACS numbers: 03.67.Lx, 32.80.Qk}

\end{abstract}
\end{minipage}

\narrowtext \vs{3.5}

%******************************************************************************
%
\section{Introduction}\label{sec-Intro}
The discrete quantum Fourier transform,
\begin{equation}\label{DFT}
    \mbox{DFT}_N:\hs{1}
    \ket{a} \mapsto
    \frac{\,1}{\hs{-0.6}\sqrt{N}}\,
    \sum_{c=0}^{N-1}
        \exp(i2\pi a c/N) \, \ket{c},
\end{equation}
which links two sets of states each labeled by integers, occurs in many
applications of quantum computing \cite{Jozsa98}. It is central to Shor's
algorithm for prime factorization \cite{Shor97}, which has applications in
public-key cryptography \cite{Rivest78}. Coppersmith \cite{Coppersmith94}
describes an efficient algorithm for implementing this transform when $N$ is a
power of $2$, achieving an exponential speed-up over the classical fast Fourier
transform (FFT) algorithm \cite{Cooley65}.

Advances in quantum FFT technology have been motivated by the difficulty of
implementing quantum logic between macroscopically distinct two-level systems,
or qubits. The difficulty arises from decoherence, the loss of coherence in a
quantum superposition due to coupling with the environment. It is known that
one-qubit gates alone, interspersed by classical measurements, suffice to build
the quantum FFT \cite{Griffiths95}, but this result is hard to implement in
practice. Approximate simulations of DFT$_N$ have also been proposed
\cite{Coppersmith94,Cleve00}, and are known to be more tolerant to phase
fluctuations in the two-qubit gates when applied in the context of Shor's
algorithm \cite{Barenco96}.

In this paper, we consider an analogue of the exact quantum FFT algorithm based
on multi-valued quantum logic \cite{Muthukrishnan00}, and propose a novel
realization of DFT$_N$ in multilevel atomic systems using wave-packet control
methods. The advantage of using $d \gthanb 2$ comput- \linebreak ational levels
in each atom is that the number of atoms needed for the algorithm is reduced by
a factor \linebreak of $\mbox{log}_2 d$. For example, $d \equal 8$ levels
stores three qubits \linebreak of information in each atom, requiring only
$Q/3$ atoms for computing DFT$_{N}$ for $N \equalb 2^Q$. Since fewer atoms
\linebreak are needed, the multilevel approach minimizes the \linebreak

\begin{minipage}{58ex}\vs{28.8}
\end{minipage}

\noindent decoherence associated with the macroscopic entanglement of these
atoms, and enables a scale-up in the implementation of the quantum FFT.

In section~\ref{sec-FFT}, we show that the elementary operations needed for the
algorithm are a Fourier transform of the $d$ levels in each atom, DFT$_d$, and
a phase gate that couples two atoms together. The $d$-level Fourier transform
takes the place of the Walsh-Hadamaard transform, which plays a prominent role
in binary quantum computation \cite{Bowden00}. The phase gate involves a
conditional coupling between two entangled atoms, and is more susceptible to
decoherence in implementation. As the number of phase gates in the quantum FFT
scales as the square of the number of atoms, the reduction in the latter in a
multilevel implementation is advantageous from a coherence-time standpoint.

We propose to implement DFT$_d$ in the atom by a change of computational basis,
as described in section~\ref{sec-Wavepacket}. The Fourier transform occurs
naturally in quantum mechanics in relating complementary representations and we
show that this can be useful for computational purposes. A dual Fourier basis
for atomic energy levels consists of localised electron wave packets at {\em
discrete times} in one Kepler orbit about the nucleus \cite{Muthukrishnan01a}.
Individual elements in the wave-packet basis can be addressed by short laser
pulses that interact with the electron when it is near the atomic core. A
change of basis from energy levels to wave packets effectively accomplishes
DFT$_d$ in the atom.

The quantum FFT then reduces to a sequence of controlled phase gates between
two atoms in {\em hybrid} bases, evolving the phases of wave-packet states in
one atom conditional on energy levels in the other. In
section~\ref{sec-Iontrap}, we consider a method for implementing such a gate in
the linear ion-trap quantum logic scheme proposed by Cirac and Zoller
\cite{Cirac95}. A multilevel phase-gate protocol in this scheme involves a
sequence of laser pulses applied to two ions in the trap.

%******************************************************************************
\section{Multi-valued quantum FFT}\label{sec-FFT}

In a system of $\hs{0.1}Q \equal\hs{0.1} \log_2 N$ qubits, DFT$_N$ can be cons-
\linebreak tructed using only two kinds of binary gates
\cite{Coppersmith94,Ekert96}. These are the single-qubit Walsh-Hadamaard
transform
\begin{equation}\label{A2}
    A_m = \frac{1}{\sqrt{2}}
        \left[
            \begin{array}{cc}
            1 & \hs{1.3}1 \\*
            1 & -1
            \end{array}
        \right],
\end{equation}
acting on qubit $m$, and the two-qubit controlled phase gate
\begin{equation}\label{B2}
    B_{lm} =
        \left[
            \begin{array}{cccc}
            1 & 0 & 0 & 0 \\*
            0 & 1 & 0 & 0 \\*
            0 & 0 & 1 & 0 \\*
            0 & 0 & 0 & e^{i\phi} \hs{-0.3}
            \end{array}
        \right],
\end{equation}
acting on qubits $l$ and $m$, where $\phi = \pi/2^{m-l}$. Specifically, it can
be shown that except for a reversal of bits in the final output,
\begin{eqnarray}\label{DFTbinary}
    \mbox{DFT}_{N} & = & (A_{Q-1}B_{Q-2,Q-1})
    (A_{Q-2}B_{Q-3,Q-1}B_{Q-3,Q-2})
    \nonumber \\*[0.3ex]
    & & \ldots (A_{1}B_{0,Q-1}B_{0,Q-2} \ldots B_{0,1}) A_0,
\end{eqnarray}
where the sequence of gates on the right side is applied from left to right.
The total number of gates is $Q(Q+1)/2 = \cO[Q^2]$, so this is an efficient
process.

To describe the multi-valued quantum FFT, we generalize the gates $A_m$ and
$B_{lm}$ to multilevel systems. Each $d$-level system is referred to as a {\em
qudit}. The states \ket{a} and \ket{c} in Eq.~(\ref{DFT}) can be written as a
tensor product of $q = \log_d N$ qudits,
\begin{eqnarray}\label{ad}
    & \ket{a} = \ketn{a_{q-1}}{a_{q-2}}{a_0}, &
\\*[0.3ex] \nonumber
        & a_m = 0,1,\dots, d-1 \mbox{ for all } m, &
\end{eqnarray}
and similarly for \ket{c}. The numbers $a_m$ represent the digits of $a$ in
base $d$. The number of qudits $q$ in the tensor product is less than the
number of qubits $Q$ by a factor of $\log_2 d$,
\begin{equation}\label{quditscaling}
    q = \log_d N = \frac{\log_2 N}{\log_2 d} = \frac{Q}{\log_2 d},
\end{equation}
which reduces the number of atoms needed for the algorithm. The multi-valued
analogue of the Walsh-Hadamaard transform $A_m$ is a $d$-level Fourier
transform,
\begin{equation}\label{Ad}
    \cA_m = \mbox{DFT}_d: \hs{0.7}
    \ket{a_m} \mapsto
    \frac{\,1}{\hs{-0.6}\sqrt{d}}\,
    \sum_{b_m = 0}^{d-1}
        \exp(i2\pi a_m b_m /d) \,\, \ket{b_m},
\end{equation}
which mixes the $d$ states in the $m$th qudit, $\ket{0},\ket{1}$, $
\ldots,\ket{d \minus 1}$, with phases determined by the Fourier kernel. The
phase gate $B_{lm}$ generalizes to the two-qudit gate
\begin{equation}\label{Bd}
    \cB_{lm}: \hs{0.5}
    \ket{a_l,b_m} \mapsto \hs{0.3}
    \exp(i2\pi a_l b_m / d^{m-l+1}) \hs{0.7} \ket{a_l,b_m},
\end{equation}
which is a diagonal transformation that advances the phase of each of the $d^2$
states in the two-qudit basis, $\ket{0,0},\ket{0,1},\ldots,\ket{d \minus 1,d
\minus 1}$, by an amount determined by the values of both qudits, $a_l$ and
$b_m$. For $d \equal 2$ and $a_l, b_m = 0 \mbox{ or } 1$ in Eqs.~(\ref{Ad}) and
(\ref{Bd}), we recover the binary gates in Eqs.~(\ref{A2}) and (\ref{B2})
respectively.

Given the above definitions for $\cA_m$ and $\cB_{lm}$, we show that a sequence
of gates similar to that in Eq.~(\ref{DFTbinary}) simulates DFT$_N$ on an
$q$-qudit register. In the multi-valued case, for $N = d^q$,
\begin{eqnarray}\label{DFTmultivalued}
    \mbox{DFT}_{N} & = & (\cA_{q-1}\cB_{q-2,q-1})
    (\cA_{q-2}\cB_{q-3,q-1}\cB_{q-3,q-2})
    \nonumber \\*[0.3ex]
    & & \ldots (\cA_{1}\cB_{0,q-1}\cB_{0,q-2} \ldots \cB_{0,1}) \cA_0.
\end{eqnarray}
Based on an argument given by Shor \cite{Shor97} for the binary quantum FFT, we
consider the matrix element of DFT$_N$ between two arbitrary states \ket{a} and
\ket{c},
\begin{equation}\label{matrixelem}
    \matrixelem{c\hs{0.1}}{\mbox{DFT}_N}{a}
    = \frac{1}{\sqrt{N}} \hs{0.4} \exp(i2\pi a c/N),
\end{equation}
and show that the sequence of gates in Eq.~(\ref{DFTmultivalued}) has the same
matrix element as above, but between states \ket{a} and \ket{b}, where \ket{b}
is defined as the `dit-reversed' version of \ket{c},
\begin{eqnarray}\label{bd}
    \ket{b} & = & \ketn{b_{q-1}}{b_{q-2}}{b_0},
\\*[0.3ex] \nonumber
            & = & \ketn{c_{0}}{c_{1}}{c_{q-1}}.
\end{eqnarray}
The least significant place in $b$ becomes the most significant place in $c$,
and vice versa. A reversal of qudits can be performed efficiently using
multi-valued permutation gates \cite{Muthukrishnan00}, or else we can simply
read out the final state in the reverse order.

To determine the amplitude $A \hs{0.2} e^{i\phi}$ of going from
$\ketn{a_{q-1}}{a_{q-2}}{a_0}$ to $\ketn{b_{q-1}}{b_{q-2}}{b_0}$ under the
sequence of gates in Eq.~(\ref{DFTmultivalued}), consider each set of gates
separated parenthetically in this sequence. First $\cA_m$ transforms \ket{a_m}
to \ket{b_m} in the $m$th qudit with amplitude $(1/\sqrt{d}) \exp(i2\pi a_m
b_m/d)$. This is followed by all the gates $\cB_{lm}$ for $m
\hs{-0.3}>\hs{-0.2} l$, each of which adds a phase \linebreak $2\pi a_l
b_m/d^{m-l+1}$ to \ket{a_l,b_m} without mixing states. The net modulus $A$ of
the transition amplitude between \ket{a} to \ket{b} is thus determined by the
product of the $\cA_m$ gates,
\begin{equation}\label{modulus}
    A =
    \left(\frac{1}{\sqrt{d}}\right)^q =
    \frac{1}{\sqrt{d^q}} =
    \frac{1}{\sqrt{N}}.
\end{equation}
The net phase $\phi$ can be separated into two parts, that due to the $\cA_m$
gates and that due to the $\cB_{lm}$ gates,
\begin{equation}\label{phase1}
    \phi = \sum_{m\hs{0.1}=\hs{0.1}0}^{q-1}
                2\pi \frac{a_m b_m}{d}
            \hs{0.3}+\hs{0.3}
            \sum_{l\hs{0.1}=\hs{0.1}0}^{q-1}
            \sum_{\hs{0.4}m \hs{0.1}>\hs{0.1} l}
                2\pi \frac{a_l b_m}{d^{m-l+1}}.
\end{equation}
Since the first term amounts to setting $l=m$ in the second term, we can
combine the terms by replacing $m>l$ with $m \hs{-0.1}\geq\hs{-0.1} l$ in the
second summation. From Eq.~(\ref{bd}), \linebreak

% ****************************************************
%
\begin{figure}[htp]
\postscript{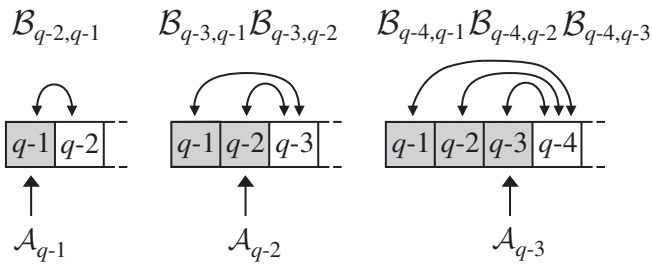}
\caption{Illustration of quantum FFT gates (based on Ref. [13]). The shaded
squares represent the qudits transformed by $\cA_m$, implemented as a change
from the energy-level to the wave-packet basis.} \vs{1.5}
\label{fig-FFTblock}
\end{figure}
%
% ****************************************************

\noindent we have $b_m \hs{-0.2}=\hs{-0.1} c_{q-1-m}$. Defining $m' \hs{-0.2}=
q \minus 1 \minus m$, the summation over $m \geq\hs{0.2} l$ becomes one over
$m'< q-l$,
\begin{equation}\label{phase4}
    \phi = \sum_{l\hs{0.1}=\hs{0.1}0}^{q-1} \sum_{\hs{0.6}m'<\hs{0.3} q-l}
            2\pi a_l c_{m'} \frac{d^{m'}}{d^{q-l}}.
\end{equation}
Including $m' \geq q \minus l$ terms in the second summation above will not
affect the phase since these extra terms are integer multiples of $2\pi$.
Hence, the two summations decouple to give
\begin{equation}\label{phase5}
    \phi = \frac{2\pi}{d^q} \hs{0.5}
            \sum_{l\hs{0.1}=\hs{0.1}0}^{q-1} a_l d^l
            \sum_{m'=0}^{q-1} c_{m'} d^{m'}
         =\, 2\pi ac/N,
\end{equation}
where we have used $N \equal d^q$ and identified $a$ and $c$ in their base-$d$
notation. From Eqs.~(\ref{modulus}) and (\ref{phase5}), we see that the net
amplitude of going from \ket{a} to \ket{b} under the sequence of gates in
Eq.~(\ref{DFTmultivalued}) is identical to that of going from \ket{a} to
\ket{c} under DFT$_N$. Thus, to within a reversal of qudits between \ket{b} and
\ket{c}, the $q(q \plus 1)/2$ gates in Eq.~(\ref{DFTmultivalued}) simulates a
quantum Fourier transform on a $q$-qudit register.

A graphical illustration of the quantum FFT is shown in
figure~\ref{fig-FFTblock}. The first three passes through the algorithm ($m = q
\minus 1, q \minus 2, q \minus 3$) are shown, corresponding to the first three
sets of gates in Eq.~(\ref{DFTmultivalued}). During each pass, an $\cA_m$ gate
first mixes the $d$ states in qudit $m$, illustrated by a shading of the
respective square in the figure, followed by a sequence of $\cB_{lm}$ gates
that couple all shaded squares with the next unshaded one. In each pass, the
$\cA_m$ gate enables a $d$-point Fourier transform that is repeated efficiently
$d^m$ times by the conditional $\cB_{lm}$ gates, achieving exponential speed-up
over the classical FFT. The speed-up is made possible by the tensor product
nature of quantum entanglement \cite{Jozsa97}. In the multi-valued case, this
corresponds to a $d$-ary tree decomposition of DFT$_N$ in terms of unitary
operations.

At the completion of the quantum FFT, each qudit is read out by measurement.
For a single multilevel system, this yields one value per qudit, corresponding
to one of the $d$ computational levels. Information stored in a superposition
of these levels is lost upon measurement, analogously to the situation in a
two-level system. The measured output of the multi-valued algorithm is thus
always a product state of the $q$ qudits, corresponding to a classical number
in base $d$.

%******************************************************************************
\section{Atomic Fourier Transform}\label{sec-Wavepacket}

Consider the implementation of the transform $\cA_m$ in an atom with $d$
computational levels. As shown in Eq.~(\ref{Ad}), this transform uniformly
mixes all the levels in the atom with phases determined by the Fourier kernel.
In the basis of energy levels, this requires precise control of the relative
phases of the levels. Such control is not feasible for large numbers of levels
in the energy basis. However, we can regard this as a problem in wave-packet
control. Atomic wave packets are superpositions of energy levels with different
phase relations among the levels. We propose to implement DFT$_d$ in the atom
by means of a dual computational basis composed of wave packets
\cite{Muthukrishnan01a}.

%**************************************************************
\subsection{Energy and wave packet bases}\label{subsec-Basis}

The Fourier transform occurs naturally in the change of state representation
from coordinate to momentum in quantum mechanics. Although time is not an
observable, we can also speak of an uncertainty relation between energy and
time. In this case, the Fourier kernel appears in the unitary time-evolution
operator, which relates the continuous dynamics of the bound atomic state to
its discrete energy spectrum. A larger number of energy levels in the
superposition leads to a greater localization in the wave packet. By
appropriately discretizing the dynamics, we can define a wave-packet basis in
the atom that is related to the energy-level basis by a discrete quantum
Fourier transform.

Consider radial wave packets \cite{Parker86}, which are superpositions of
Rydberg energy levels with low angular momentum. These levels have long
radiative life times, approaching a millisecond for principal quantum number $n
> 100$. Take $d$ energy levels centered at $\nbar$ with angular momentum
$l =\hs{-0.1} 1$ to represent the computational basis in the atom,
\begin{eqnarray}\label{energylevels}
    & \ket{j}_\nu = \ket{\nbar + j, 1, 0}, &
    \\*[0.8ex] \nonumber
    & j = -d/2+1, -d/2+2, \ldots, d/2, &
\end{eqnarray}
where the subscript $\nu$ denotes a state in the energy-level basis. We have
assumed above that $\nbar$ is an integer and $d$ is an even number for
simplicity, but the arguments below are easily extended to non-integer $\nbar$
and odd $d$. A uniform superposition of the $d$ levels corresponds to a
radially-localized wave packet in space whose time evolution is given by
\begin{equation}\label{wavepacket}
    \ket{\psi(t)} = \frac{1}{\sqrt{d}} \sum_{j}
                     \exp(-i\omega_j t) \hs{0.2}
                     \ket{j}_\nu,
\end{equation}
where $\hbar \omega_j$ is the energy of the $j$th level in the superposition,
and $\hbar \omega_0$ is the mean energy of the wave packet corresponding to
principal quantum number $\nbar$. To separate the classical and revival
dynamics of the wave packet, we expand $\omega_j -\omega_{0}$ in a Taylor
series in $j = n-\nbar$,
\begin{equation}\label{Taylor}
    \omega_j - \omega_{0} = 2\pi
        \left[\frac{j}{T_K} - \frac{j^2}{2!\hs{0.2}T_{\mathrm{rev}}} +
            \frac{j^3}{3!\hs{0.2}T_{\mathrm{sr}}} - \cdots \right].
\end{equation}
The Kepler period $T_K  \equal 2\pi\nbar^3$ (in atomic units) measures the
round-trip time for the wave packet traveling between the inner and outer
turning points of the classical orbit. This corresponds to a radial shell of
probability distribution varying periodically in size. The revival time
$T_{\mathrm{rev}}$ and the super-revival time $T_{\mathrm{sr}}$ describe
higher-order effects such as dispersion and revivals.

\begin{figure}[htp]
\postscript{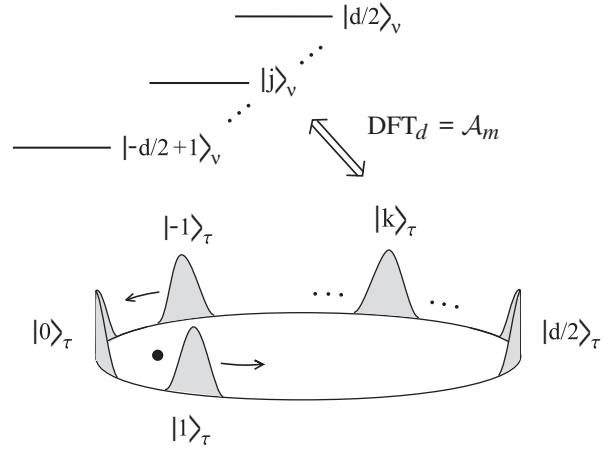}
\vs{1}\caption{The Fourier conjugate of the energy-level basis $\ket{j}_\nu$
consists of wave-packet states $\ket{k}_\tau$ evenly distributed in time around
a classical orbit. A radial wave-packet basis can be thought of as an ensemble
of such orbits with different orientations for the ellipses.}
\label{fig-WPbasis}\vs{1.3}
\end{figure}

Define an orthogonal basis of wave-packet states corresponding to $d$ discrete
times during the classical Kepler evolution of the radial wave packet
\cite{Muthukrishnan01a},
\begin{eqnarray}\label{wavepacketbasis}
    \ket{k}_{\tau}
    & =\hs{0.2} & \frac{1}{\sqrt{d}} \sum_{j}
            \exp(-i 2\pi j k /d) \hs{0.2}
            \ket{j}_{\nu}
\\*[0.5ex] \nonumber
    & \approx\hs{0.2} & \ket{\psi(kT_K/d)},
\end{eqnarray}
where $k = -d/2+1, -d/2+2, \ldots, d/2$, and the subscript $\tau$ denotes a
state in the wave-packet basis. Thus $k=0$ corresponds to the wave-packet state
centered at the inner turning point of the classical orbit, and $k \equal \pm
|k|$ correspond to wave-packet states moving in opposite directions at some
intermediate location in the orbit, as illustrated in figure~\ref{fig-WPbasis}.

The energy-level basis $\ket{j}_\nu$ is related to the wave-packet basis
$\ket{k}_\tau$ by a $d$-level Fourier transform,
\begin{eqnarray}\label{basischange1}
    \cA_m \ket{j}_\nu
    & =\hs{0.2} & \frac{1}{\sqrt{d}} \sum_{j'}
        \exp(i 2\pi jj'/d) \hs{0.2}
        \ket{j'}_\nu
\\* \nonumber
    & =\hs{0.2} & \ket{k=-j}_\tau,
\end{eqnarray}
where we have used Eqs.~(\ref{Ad}) and (\ref{wavepacketbasis}). This suggests
that $\cA_m$ can be implemented in the atom by a change of computational basis
from that of energy eigenstates to that of wave-packet states. A change of
basis does not involve any free time evolution or real-time processing in the
atom, which means that there is no computational cost to realizing the $\cA_m$
gates in this manner.

To understand the effect of the basis change on the algorithm, we refer again
to figure~\ref{fig-FFTblock}. Recall that the shading of each square in the
figure corresponds to the application of a $\cA_m$ gate to that qudit in the
array, and the following $\cB_{lm}$ gates in that pass through the algorithm
couple all the shaded squares with the next unshaded one. If $\cA_m$ is
regarded as a change of basis from energy levels to wave packets, then the
\linebreak

% ****************************************************
%
% ****************************************************

\noindent shaded squares are to be read in the wave-packet basis. Consequently,
each $\cB_{lm}$ gate involves different bases for the two qudits $l$ and $m$ in
the transformation. The quantum FFT thus reduces to a series of conditional
two-qudit phase transforms $\cB_{lm}$ performed in hybrid bases.

In the atomic case, $\cB_{lm}$ can be regarded as phase shifts applied to each
wave-packet state $\ket{k}_\tau$ in the $m$th atom conditional on each energy
eigenstate $\ket{j}_\nu$ in the $l$th atom. In view of
Eq.~(\ref{basischange1}), we rewrite Eq.~(\ref{Bd}) as
\begin{eqnarray}\label{Bd2}
    & \cB_{lm}: \hs{0.7}
        \ket{j}_\nu \ket{k}_\tau \mapsto \hs{0.3}
        \exp(i\phi_{jk}) \hs{0.3}
        \ket{j}_\nu \ket{k}_\tau, &
\\*[0.5ex] \nonumber
    & \phi_{jk} = -2\pi j k/d^{m-l+1}. &
\end{eqnarray}
We describe a protocol for implementing $\cB_{lm}$ in a linear ion trap in
section~\ref{sec-Iontrap}. This requires coherent control of the wave-packet
basis in the target atom, which we discuss below.

%**************************************************************
\subsection{Coherent wave packet control}\label{subsec-Control}

Consider an arbitrary Rydberg state in the wave-packet basis,
\begin{equation}\label{intpicture}
        \ket{\psi(t)} = \exp(-i\omega_{0}t)
         \sum_k \tilde{b}_k(t)  \hs{0.3} \ket{k}_\tau,
\end{equation}
where $\tilde{b}_k$ are slowly varying amplitudes from which we have removed
the average frequency $\omega_0$ corresponding to the mean Rydberg level
$\nbar$ in Eq.~(\ref{energylevels}). Since the wave-packet states are not
stationary, the amplitudes $\tilde{b}_k$ evolve in time. During a Kepler
period, the periodic motion of the radial wave packet corresponds to a cyclic
permutation in the amplitudes,
\begin{equation}\label{cyclicevol}
    \tilde{b}_k(mT_K/d) = \tilde{b}_{k-m}(0),
\end{equation}
where $m$ is an integer and $k-m$ is taken modulo $d$. At later times,
higher-order terms in the Taylor expansion of Eq.~(\ref{Taylor}) become
significant, and the dispersion of the Rydberg wave function mixes the
amplitudes $\tilde{b}_k$ nontrivially. However at the revival times
$T_{\mathrm{rev}}$, the wave function reforms into the original state and
nearly recovers the initial distribution of amplitudes in the wave-packet
basis.

An applied laser field interacts strongly with a Rydberg wave packet when it is
near the atomic core. This phenomenon underlies the excitation and
photo-ionization of radial wave packets \cite{Wolde88}, and we use this as a
means for coherent control of individual amplitudes $\tilde{b}_k$ in the
wave-packet basis. The idea is to use short laser pulses to transfer a chosen
amplitude to the ground state for time-resolved processing.

Consider a broadband laser pulse with a spectral width $\sim d/T_K$ that
couples all the Rydberg levels in Eq.~(\ref{energylevels}) to the ground state
\ket{g} in the atom. If the pulse is transform-limited, it has a temporal width
less than $T_K/d$, and we can to good approximation ignore the free Kepler
evolution of the wave-packet amplitudes shown in Eq.~(\ref{cyclicevol}). Only
the amplitude $\tilde{b}_0$ changes significantly during the pulse,
corresponding to the wave-packet state $\ket{0}_\tau$ nearest the atomic core.
This state undergoes Rabi oscillations with the ground state
\cite{Muthukrishnan01a},
\begin{eqnarray}
\label{bemeqtwolevel}
    \dot{b}_g & \cong & \frac{i}{2} \hs{0.2} f(t) \hs{0.2} \tilde{\Omega}_0
        \exp(-i\Delta_{0} t) \hs{0.2} \tilde{b}_0,
\\*[1ex] \label{bmeqtwolevel}
    \dot{\tilde{b}}_0 & \cong & \frac{i}{2} \hs{0.2} f(t) \hs{0.2} \tilde{\Omega}_0
        \exp(i\Delta_{0} t) \hs{0.2} b_g.
\end{eqnarray}
where $f(t)$ is the pulse profile, $\Delta_{0} = \omega_{0} \minus \omega$ is
the de-\linebreak tuning of the center frequency $\omega$ of the pulse from the
average Rydberg frequency, and $\tilde{\Omega}_0$ is proportional to the
average of the Rabi frequencies $\Omega_{gj}$ for the ground-to-Rydberg
transitions,
\begin{equation}\label{Rabim}
    \tilde{\Omega}_0 = \frac{1}{\sqrt{d}} \sum_{j} \Omega_{gj}.
\end{equation}
A localized wave packet behaves classically for $t \sim T_K$, and the strong
coupling to the laser field near the core can be understood as a large momentum
transfer to the electron near the nucleus. The rate at which energy is absorbed
from the field ${\bf E}$ is proportional to ${\bf p}\cdot{\bf E}$, and the
electron momentum ${\bf p}$ is maximum at the inner turning point.

The two-level system of Eqs.~(\ref{bemeqtwolevel}) and (\ref{bmeqtwolevel})
allows selective wave-packet processing in the atom. In particular, it allows
phase control of individual wave-packet states in the Rydberg basis, as needed
to implement $\cB_{lm}$. To see this, note that a $\pi$-pulse of duration less
than $T_K/d$ de-excites only that part of the Rydberg wave function associated
with a {\em single} wave-packet amplitude $\tilde{b}_k$, creating a `dark' wave
packet in its place in the Rydberg basis. \linebreak

% ****************************************************
\begin{figure}[htp]
\postscript{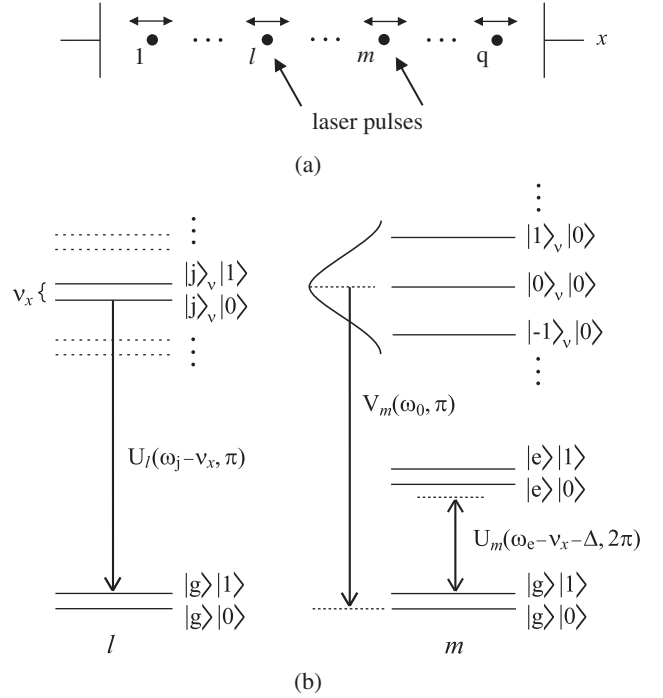}
\vs{1}\caption{Phase-gate implementation. (a) Ion trap with pulses applied to
ions $l$ (control) and $m$ (target). (b) Level scheme in each ion, with Rydberg
levels $j$ and auxiliary levels $g$ and $e$. Three types of pulses used:
$U_l(\omega_j -\nu_x, \pi)$, $V_m(\omega_{0},\pi)$, and $U_m(\omega_e \minus
\nu_x \minus \Delta, 2\pi)$. See text for details.}
\label{fig-PGiontrap}\vs{1.5}
\end{figure}
%
% ****************************************************

\noindent The phase of $\tilde{b}_k$ in the ground state can be controlled by a
narrow-band $2\pi$-pulse that couples this state to an auxiliary energy level
in the atom (level $e$ in figure~\ref{fig-PGiontrap}). The phase-shifted
amplitude can be restored to the Rydberg basis at a time commensurate with when
the dark wave-packet state returns to the atomic core. This accomplishes a
phase shift of a selected wave-packet state in the target atom.

Since the wave-packet basis is not stationary, the time between successive
interactions with the Rydberg basis is dictated by the free atomic time scales.
This time interval has to an integer number of Kepler periods during which the
Rydberg state has not incurred much dispersion, or alternately, when the free
time evolution undergoes a revival in the \Sch picture. This ensures that the
decomposition of the Rydberg wave function in the wave-packet basis has not
changed appreciably between pulses.

%******************************************************************************
\section{Phase Gate in the Linear Ion Trap}\label{sec-Iontrap}

Consider an implementation of the two-qudit phase gate $\cB_{lm}$ in the linear
ion-trap scheme for quantum computing \cite{Cirac95}. Assuming that $d$ energy
levels in each trapped ion represent a qudit, consider a series of laser pulses
applied to ions $l$ and $m$ in the trap, as illustrated in
figure~\ref{fig-PGiontrap}(a). Our goal is to control the phases of the
wave-packet states $\ket{k}_\tau$ in the $m$th ion conditional on the energy
eigenstates $\ket{j}_\nu$ in the $l$th ion, as required by Eq.~(\ref{Bd2}).

The ions are assumed to be in the vibrational ground state and oscillate
synchronously in the center-of-mass normal mode in the trap. Assuming that the
interaction field has a standing-wave pattern along the trap axis, two kinds of
interactions have been proposed in this scheme, labeled $U$ and $V$
\cite{Cirac95}. The $V$ interaction arises when the ion is at the antinode of
the standing wave, and the laser resonantly couples two internal states in the
ion according to the unitary evolution operator
\begin{equation}
\label{hamiltonianV}
    \hat{V}(t) =
    \exp\hs{-0.4}\left[ it \hs{0.2} \frac{\Omega}{2} \hs{0.2}
    (\hat{\sigma}^{\dagger} + \hat{\sigma})
    \right],
\end{equation}
where $\Omega$ is the Rabi frequency and $\hat{\sigma}$ is the lowering
operator for the atomic transition. Alternately, the $U$ interaction arises
when the ion is at the node of the standing wave and the laser is detuned off
resonance to an atomic transition by the trap frequency $\nu_x$. We consider
the lowest two trap states, \ket{0} and \ket{1}. For atomic levels $g$ and $e$,
where $e$ is the upper level, we find that the states \ket{g}\ket{1} and
\ket{e}\ket{0} are coupled by the unitary operator
\begin{equation}
\label{hamiltonianU-}
    \hat{U}(t) =
    \exp\hs{-0.4}\left[-it \hs{0.2} \frac{\eta}{\sqrt{q}} \frac{\Omega}{2}\hs{0.2}
    (\hat{\sigma}^{\dagger} \hat{a}
    + \hat{\sigma} \hat{a}^{\dagger})
    \right],
\end{equation}
where $\hat{a}$ is the trap lowering operator and $q$ is the number of ions in
the trap. The Lamb-Dicke parameter $\eta$ is defined as
\begin{equation}\label{LDP}
    \eta = k_{x} \sqrt{\frac{\hbar}{2m\nu_x}},
\end{equation}
where $m$ is the mass of each ion, and $k_{x}$ is the wave vector along the
trap axis. The unitary evolution in Eq.~(\ref{hamiltonianU-}) is valid in the
limit that $\eta \ll 1$.

Consider the two-ion Rydberg wave function at some time $t_0$ when the trap has
been initialized to \ket{0},
\begin{equation}\label{wave0full}
    \ket{\Psi(t_0)} = \sum_{j'} \sum_{k'} c_{j'k'}(t_0) \hs{0.2} \ket{j',k'}\ket{0},
\end{equation}
where we use the abbreviation $\ket{j'}_\nu \ket{k'}_\tau \equal\hs{0.1}
\ket{j',k'}$, and the summations over $j'$ and $k'$ run over the $d$ components
of the energy-level and wave-packet bases in ions $l$ and $m$ respectively. The
coefficients $c_{j'k'}$ are the \Sch picture amplitudes whose free time
evolution has two contributions, one due to the phase evolution of the energy
levels in the $l$th ion, and another due to the periodic evolution of the
wave-packet amplitudes in the $m$th ion. We have to keep these contributions in
mind as we pursue a phase gate in the hybrid basis.

When the $k$th wave-packet element in the $m$th ion is near the atomic core,
the methods described in section~\ref{subsec-Control} can be used to transfer
the corresponding amplitudes $c_{j'k}$ to the ground state \ket{g} in the ion.
This is done by applying a broadband $\pi$-pulse of the $V$ type, denoted by
$V_m(\omega_{0},\pi)$ in figure~\ref{fig-PGiontrap}(b). The pulse spectrum is
centered on the mean frequency $\omega_{0}$ and has a duration less than
$T_K/d$ that is an integer multiple of $\pi/\tilde{\Omega}_0$. This only
affects the wave-packet state nearest the atomic core, denoted by $k' = k$, and
leaves the two ions in the state
\begin{eqnarray}
    \lefteqn{\ket{\Psi(t_1)}}
\nonumber \\* \nonumber
        & & \mbox{ } = \sum_{j'} \left[
        c_{j'k}(t_1) \hs{0.2} \ket{j',g}\ket{0} + \hs{-0.4}
        \sum_{k'\neq k} c_{j'k'}(t_1) \hs{0.2} \ket{j',k'}\ket{0}
        \right]
\\*[0.5ex] \label{wave1full}
        & & \mbox{ } = \ket{\Psi_k(t_1)} +
        \sum_{j'} \sum_{k'\neq k} c_{j'k'}(t_1) \hs{0.2} \ket{j',k'}\ket{0}.
\end{eqnarray}
The second term in Eq.~(\ref{wave1full}) represents that part of the wave
function in the $m$th ion that is left in the Rydberg manifold, and we suppress
this term briefly. The first term corresponds to the $k$th wave-packet state
that has been de-excited, which can be written as
\begin{eqnarray}
    \ket{\Psi_k(t_1)} & = & \sum_{j'}
        c_{j'k}(t_1)\hs{0.1} \ket{j',g}\ket{0}
\nonumber \\* \label{wave1}
    & = & c_{jk}(t_1)\hs{0.1} \ket{j,g}\ket{0} +
        \sum_{j'\neq j} c_{j'k}(t_1)\hs{0.1} \ket{j',g}\ket{0},
\end{eqnarray}
where a particular energy level $j$ is taken out of the $j'$ summation. We seek
to de-excite this level to the ground state in the $l$th ion, conditional on
exciting the trap. This is done by applying a narrow-band $\pi$-pulse of the
$U$ type to the $l$th ion, which has a pulse duration that is an integer
multiple of $\pi /(\eta \Omega_{gj}/\sqrt{q})$. The laser frequency is tuned to
$\omega_j$ for the $j$th Rydberg level. This pulse is labeled $U_l(\omega_j
\minus \nu_x,\pi)$ in figure~\ref{fig-PGiontrap}(b), and transforms
\ket{\Psi_k(t_1)} to
\begin{equation}\label{wave2}
    \ket{\Psi_k(t_2)} =
    c_{jk}(t_2)\hs{0.2} \ket{g,g}\ket{1} +
        \sum_{j'\neq j} \hs{-0.2} c_{j'k}(t_2)\hs{0.2} \ket{j',g}\ket{0},
\end{equation}
where the coefficients have evolved in phase from $t_1$ to $t_2$ due to the
free time evolution of the energy levels in the \Sch picture.
Equation~(\ref{wave2}) shows that the trap is excited only when both ions are
in the ground state \ket{g,g}, corresponding to the initial state \ket{j,k} at
time $t_0$. Hence, the $U$ pulse has created entanglement between the trap
state and the internal states of the two ions.

To implement $\cB_{lm}$, we need to shift the phase of \ket{j,k} by $\phi_{jk}$
according to Eq.~(\ref{Bd2}). In state \ket{\Psi_k(t_2)}, this corresponds to
evolving the phase of \ket{g,g}\ket{1} by $\phi_{jk}$ without affecting the
other basis states. To do this, consider the auxiliary level $e$ in the $m$th
ion shown in figure~\ref{fig-PGiontrap}(b). Applying a $U$ pulse of $2\pi$
duration couples states \ket{g}\ket{1} and \ket{e}\ket{0} in the $m$th ion. For
a laser detuning of $\Delta$, the interaction phase of \ket{g,g}\ket{1} evolves
by an integer multiple of $\pi(1+\Delta/\Omega_{ge})$, which can be controlled
to achieve $\phi_{jk}$. This pulse is denoted $U_m(\omega_e \minus\hs{0.1}
\nu_x \hs{-0.3}-\hs{-0.3} \Delta, 2\pi)$ in the figure, and transforms
\ket{\Psi_k(t_2)} to
\begin{eqnarray}
    \ket{\Psi_k(t_3)} & = &
    c_{jk}(t_3) \exp(i\phi_{jk}) \hs{0.2} \ket{g,g}\ket{1}
\nonumber \\*[0.5ex] \label{wave3}
    & & + \sum_{j'\neq j} c_{j'k}(t_3) \hs{0.2} \ket{j',g}\ket{0},
\end{eqnarray}
giving a controlled phase shift $\phi_{jk}$ to the state \ket{g,g}\ket{1} as
desired. We now reverse the operation that took us from \ket{\Psi_k(t_1)} to
\ket{\Psi_k(t_2)} by applying $U_l(\omega_j \hs{-0.1}-\nu_x,\pi)$ again to the
$l$th ion, creating
\begin{eqnarray}
    \ket{\Psi_k(t_4)} & = &
    c_{jk}(t_4) \exp(i\phi_{jk}) \hs{0.2} \ket{j,g}\ket{0}
\nonumber \\*[0.5ex] \label{wave4}
    & & + \sum_{j'\neq j} c_{j'k}(t_4) \hs{0.2} \ket{j',g}\ket{0}.
\end{eqnarray}

Lastly, the $m$th ion state \ket{g} is restored to $\ket{k}_\tau$ by applying
$V_m(\omega_{0},\pi)$ again at a time that is commensurate with when the `dark'
radial wave-packet element corresponding to $k$ returns to the atomic core.
Since this is a $V$ pulse, it does not affect the trap. The resulting state is
\begin{eqnarray}
    \ket{\Psi_k(t_5)} & = &
    c_{jk}(t_5) \exp(i\phi_{jk}) \hs{0.2} \ket{j,k}\ket{0}
\nonumber \\*[0.5ex] \label{wave5}
    & & + \sum_{j'\neq j} c_{j'k}(t_5) \hs{0.2} \ket{j',k}\ket{0}.
\end{eqnarray}
The $k' \neq k$ terms in Eq.~(\ref{wave1full}) are unaffected by the
combination of the five pulses used above. Including their contribution to the
final wave function, we get
\begin{eqnarray}
    \ket{\Psi(t_5)} & = &
    c_{jk}(t_5)\hs{0.1} \exp(i\phi_{jk}) \hs{0.2} \ket{j,k}\ket{0}
\nonumber \\*[0.5ex] \label{wave5full}
    & & + \sum_{j'\neq j} \sum_{k' \neq k}
        c_{j'k'}(t_5)\hs{0.1} \ket{j',k'}\ket{0}.
\end{eqnarray}
Comparing Eqs.~(\ref{wave0full}) and (\ref{wave5full}), we see that the
sequence of five pulses,
\begin{eqnarray}
& V_m(\omega_{0},\pi) \hs{0.5}
U_l(\omega_j \minus \nu_x,\pi) \hs{0.5}
U_m(\omega_e \minus \nu_x \minus \Delta, 2\pi) &
\nonumber \\*[0.3ex]
& U_l(\omega_j \minus \nu_x,\pi) \hs{0.5}
V_m(\omega_{0},\pi), &
\end{eqnarray}
accomplishes a controlled phase shift of the hybrid state $\ket{j,k} = \ket{j}_\nu
\ket{k}_\tau$, as desired. This procedure is repeated for each of the $d^2$
states in the two ions, leading to the phase gate $\cB_{lm}$ in mixed
energy-level and wave-packet bases.

The coefficients $c_{j'k'}(t_5)$ are different from $c_{j'k'}(t_0)$ due to the
phase evolution of the energy levels in both ions during the time interval $t_5
\minus t_0$. This is also responsible for the non-stationarity of the
wave-packet basis, which makes this procedure sensitive to when the $k$th
wave-packet amplitude in the $m$th ion is de-excited from, and excited to, the
Rydberg manifold. The time interval between the two $V_m$ pulses in the
sequence is determined by the free atomic time scales governing the classical
or revival dynamics of the wave packets. By appropriate timing of these pulses,
we can implement a conditional phase-shift between the two atoms.

%******************************************************************************
\section{Conclusion}\label{sec-Conclusion}

This paper shows that the quantum FFT can be simplified using a multilevel
wave-packet approach to its implementation. In atomic systems, the basic logic
gate is a multilevel Fourier transform $\cA_m$ in each atom, which is
equivalent to a change of basis from energy levels to wave packets. Such an
implementation takes advantage of the natural Fourier transform relation
between energy and time in quantum mechanics. The FFT then reduces to a series
of two-qudit phase gates $\cB_{lm}$ in hybrid bases, which we have considered
in the context of the linear ion-trap scheme for quantum computing.

The advantage of the multilevel approach is a reduction in the number of
entangled quantum systems (e.g. trapped ions) by a factor of $\log_2 d$
compared to the binary case. For the same reason, the number of logic gates
needed to simulate DFT$_N$ is fewer by a factor of $(\log_2 d)^2$, as seen by
comparing Eqs.~(\ref{DFTbinary}) and (\ref{DFTmultivalued}). However, this
comes at the cost of larger elementary gates, up to $d^2$-dimensional in the
case of $\cB_{lm}$. The trade-off in computation time depends very much on the
particular implementation scheme used, which dictates the physical time taken
to perform each gate.

The bottleneck for the gate operation time in the linear ion-trap scheme is the
trap frequency $\nu_x$ \cite{Jonathan00}, typically kHz to MHz, which limits
the speed of the narrow-band $U$ pulses. This is due to the need for selective
entanglement between the internal state of the ion and its motional state in
the trap. By comparison, the Rydberg atomic time scales are much faster,
typically in the ns to $\mu$s range, allowing much faster execution times in
principle for the phase-gate protocol given in section~\ref{sec-Iontrap}.

There are two advantages to a wave-packet approach to multilevel processing.
One is the quantum Fourier transform itself, which is integral to the approach
and key to quantum computing. The other is the feasibility of the control
scheme for atomic systems. Universal control of multiple energy levels in the
atom requires multiple lasers tuned to the neighbouring transitions
\cite{Muthukrishnan00}. This is much easier in the time domain, where
wave-packet transforms can be achieved by controlling the timing and durations
of a sequence of laser pulses. The implementation of the quantum FFT in
multilevel systems is thus made more feasible using wave-packet methods.

This work was supported by the Army Research Office through the MURI Center for
Quantum Information.

\vs{-0.5}
%******************************************************************************
\makeatletter


\begin{references}

\vs{-8.2}

\bibitem[*]{email}Electronic Address: amuthuk@optics.rochester.edu \\[-1ex]

\bibitem{Jozsa98}
R.~Jozsa, Proc. Roy. Soc. Lond. A {\bf 454}, 323 (1998).

\bibitem{Shor97}
P.~W.~Shor, SIAM J. Comp. {\bf 26}, 1484 (1997).

\bibitem{Rivest78}
R.~L.~Rivest, A.~Shamir and L.~Adleman, Commun. of A.C.M. {\bf 21}, 120 (1978).

\bibitem{Coppersmith94}
D.~Coppersmith, IBM Res. Report RC 19642, T.J. Watson Research Center, Yorktown
Heights, NY, 1994.

\bibitem{Cooley65}
J. W. Cooley and J. W. Tukey, Math. Computation {\bf 19}, 297 (1965).

\bibitem{Griffiths95}
R.~B.~Griffiths and C.~Niu, Phys. Rev. Lett. {\bf 76}, 3228 (1996).

\bibitem{Cleve00}
R.~Cleve and J.~Watrous, in {\em Proc. 41st Annual Symp. Found. Comp. Sci.}
(IEEE, Los Alamitos, CA, 2000), p. 526. Also at quant-ph/0006004.

\bibitem{Barenco96}
A. Barenco et al., Phys. Rev. A {\bf 54}, 139 (1996).

\bibitem{Muthukrishnan00}
A.~Muthukrishnan and C.~R.~Stroud, Jr., Phys. Rev. A {\bf 62}, 52309 (2000).

\bibitem{Bowden00}
Bowden,~C.~M., Chen,~G., Diao,~Z., and Klappenec-\linebreak ker,~A.,
quant-ph/0007122 at xxx.lanl.gov.

\bibitem{Muthukrishnan01a}
Muthukrishnan,~A., and Stroud,~Jr.,~C.~R., quant-ph/0106165 at xxx.lanl.gov.

\bibitem{Cirac95}
J.~I. Cirac and P. Zoller, Phys. Rev. Lett. {\bf 74}, 4091 (1995).

\bibitem{Ekert96}
A.~Ekert and R.~Josza, Rev. Mod. Phys. {\bf 68}, 733 (1996).

\bibitem{Jozsa97}
R.~Jozsa, to appear in {\em Geometric Issues in the Foundations of Science},
eds. S. Huggett et al. (Oxford Univ. Press). Also at quant-ph/9707034.

\bibitem{Parker86}
J.~Parker and C.~R.~Stroud,~Jr., Phys. Rev. Lett. {\bf 56}, 716 (1986).

\bibitem{Wolde88}
A.~ten Wolde et al., Phys. Rev. Lett. {\bf 61}, 2099 (1988).

\bibitem{Jonathan00}
D.~Jonathan, M.~B.~Plenio and P.~L.~Knight, Phys. Rev. A {\bf 62}, 42307
(2000).

\end{references}
\end{document}